\documentclass[aps,prl,twocolumn,superscriptaddress,showpacs,preprintnumbers]{revtex4}
\usepackage{graphicx}
\usepackage{amssymb}

\newcommand{\slrr}      {$T_1^{-1}$}

\newcommand{\afc}       {antiferromagnetic}

\newcommand{\cecoin}    {CeCoIn$_5$}
\newcommand{\cemin}    {CeMIn$_5$}
\newcommand{\cecoincdx}   {CeCo(In$_{1-x}$Cd$_{x})_5$}
\newcommand{\cecoincdten}   {CeCo(In$_{0.9}$Cd$_{0.1}$)$_5$}

\newcommand{\cerhin}    {CeRhIn$_5$}

\newcommand{\tc}     {$T_{\rm c}$}

\newcommand{\slrrtext}  {spin lattice relaxation rate}

\begin{document}


\preprint{LA-UR-06-5450}

\title{Interacting Antiferromagnetic Droplets in Quantum Critical \cecoin}
\author{R. R. Urbano}
\affiliation{Condensed Matter and Thermal Physics, Los Alamos
National Laboratory, Los Alamos, NM 87545, USA}
\author{B.-L. Young}
\affiliation{Department of Electrophysics, National Chiao Tung
University, Hsinchu 300, Taiwan}
\author{N. J. Curro}
\author{J. D. Thompson}
\affiliation{Condensed Matter and Thermal Physics, Los Alamos
National Laboratory, Los Alamos, NM 87545, USA}
\author{L. D. Pham}
\affiliation{University of California, Davis, CA 95616}
\author{Z. Fisk}
\affiliation{University of California, Irvine, CA 92697-4573}

\date{\today}

\begin{abstract}

The heavy fermion superconductor CeCoIn$_5$ can be tuned between
superconducting and antiferromagnetic ground states by hole doping
with Cd.   Nuclear magnetic resonance (NMR) data indicate that these
two orders coexist microscopically with an ordered moment $\sim$0.7
$\mu_{\rm B}$.  As the ground state evolves, there is no change in
the low frequency spin dynamics in the disordered state.  These
results suggest that the magnetism emerges locally in the vicinity
of the Cd dopants.

\end{abstract}


\pacs{71.27.+a, 76.60.-k, 74.70.Tx, 75.20.Hr}

\maketitle

The discovery of superconductivity in the layered \cemin\ system has
reignited interest in the low temperature physics of the Kondo
lattice.  The \cemin\ materials, with M = Rh, Co, or Ir, exhibit
antiferromagnetism, superconductivity or the coexistence of these
two orders depending on the external hydrostatic pressure and the
particular alloy content of the M element \cite{jdtreview}. \cecoin\
is particularly interesting as it has the highest \tc\ for a
Ce-based heavy fermion superconductor, and the normal state exhibits
non-Fermi liquid behavior that may be associated with a quantum
critical point (QCP) \cite{romanQCPCeCoIn5}. Recently Pham and
coworkers discovered that the ground state of \cecoin\ can be tuned
reversibly between superconducting and antiferromagnetic by
substituting Cd for In, with a range of coexistence for intermediate
dopings \cite{FiskCddoping}.  Although other materials exhibit
coexisting antiferromagnetism and superconductivity, the \cemin\
system is unique in that it can be continuously tuned by hole
doping.

\begin{figure}
\includegraphics[angle=0,width=\linewidth]{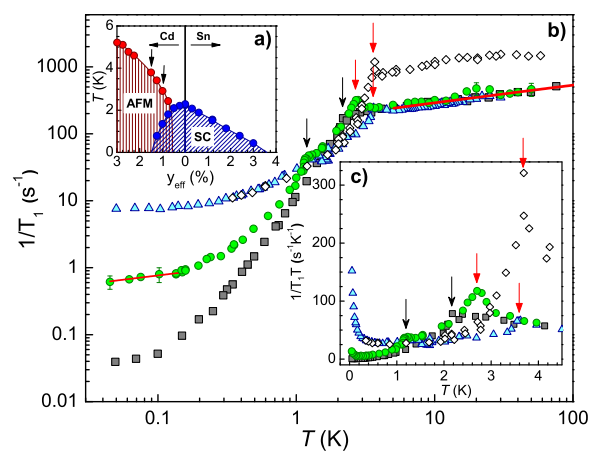}
\caption{\label{fig:T1vsT} (a) $T_c$ (blue) and $T_N$ (red) versus
nominal doping in CeCo(In$_{1-x}$Cd$_x$)$_5$ and
CeCo(In$_{1-x}$Sn$_x$)$_5$ \cite{FiskCddoping,ronningbauer}.  Arrows
show the concentrations reported in this study, and crosses show the
points corresponding to the spectra in Fig. \ref{fig:spectra}. (b)
\slrr\ versus temperature in \cecoincdx\ with $x=0.15$ (blue,
$\blacktriangle$), $x=0.10$ (green, $\bullet$), $x=0$ (grey
$\blacksquare$, from \cite{kohoriCeCoIn5QCP}) and \cerhin\
($\lozenge$). $T_N(x)$ is indicated by red arrows and $T_c(x)$ by
black arrows. (c) $(T_1T)^{-1}$ versus temperature.}
\end{figure}

 In the Doniach model of a Kondo lattice, localized spins
interact with conduction electron spins via an exchange interaction,
$J$, and the ground state depends sensitively on the product
$JN(0)$, where $N(0)$ is the density of conduction electron states
at the Fermi level \cite{doniach}. For $JN(0)\gg 1$, Kondo screening
of the local moments by the conduction electron spins dominates,
resulting in a spin liquid ground state.  At the other extreme where
$JN(0)\ll 1$, the indirect exchange (RKKY) between local spins
mediated by the conduction electrons dominates, and the ground state
is antiferromagnetic. When $JN(0)\sim 1$  there is a QCP where
$T_N\rightarrow 0$. Superconductivity typically emerges in this
regime where competing interactions lead to complex behavior and
small perturbations can drastically alter the ground state. In
\cecoin\ Cd doping introduces holes that may modify the Fermi
surface(s) of the conduction electrons, changing $N(0)$ and hence
the ground state.  As the Fermi surface evolves with hole doping,
either superconductivity or antiferromagnetism may emerge depending
on the quantity $JN(0)$ in much the same way that pressure changes
the ground state by modifying $J$ \cite{jdtreview}. An alternative
to this global interpretation is that the Cd acts as a local defect
that nucleates antiferromagnetism in a quantum critical system
\cite{castronetolocaldefect,schmalianlocaldefect}.  In a system
close to an
 QCP, a local perturbation can induce droplets of local antiferromagnetic order.
As the correlation length grows and reaches the scale of the
distance between impurities, the system can undergo long-range
order.

In this Letter we report NMR data in \cecoincdx\ that supports the
latter scenario. We report data for $x=0.10$ and $x=0.15$, where $x$
is the nominal concentration as reported in Ref. \cite{FiskCddoping}
(Fig. \ref{fig:T1vsT}a). At $x=0.10$, the superconducting transition
has been suppressed from 2.3 K to $T_{\rm c}=1.2$ K, and
antiferromagnetism emerges at $T_N = 2.8$ K; for $x=0.15$, $T_{\rm
c}=0$ K and $T_N=3.7$ K. We find that the \slrrtext, \slrr, falls at
both $T_N$ and \tc, in agreement with the thermodynamically
determined values \cite{FiskCddoping}, and the spectra reveal a
homogeneous internal field below $T_N$. These results imply that
both the antiferromagnetism and superconductivity coexist
microscopically throughout the bulk of the material. In the
disordered state, the NMR spectra are inhomogeneously broadened and
the second moment of the Co resonance is strongly temperature
dependent, suggesting that the electronic response of the sample
depends on the proximity to a doped Cd site. The Knight shift, $K$,
is nearly identical to pure \cecoin, with a strong anomaly at
$T^*\sim 60$ K, and \slrr\ exhibits an unusual $\sim T^{1/4}$
variation of \slrr\ that is doping independent above 5 K. These
results are surprising, as they indicate that the low energy spin
dynamics are nearly independent of the ground state order, and that
the fate of the 4f degrees of freedom are determined only below
$T\approx2$\tc.

\begin{figure}
\includegraphics[width=\linewidth]{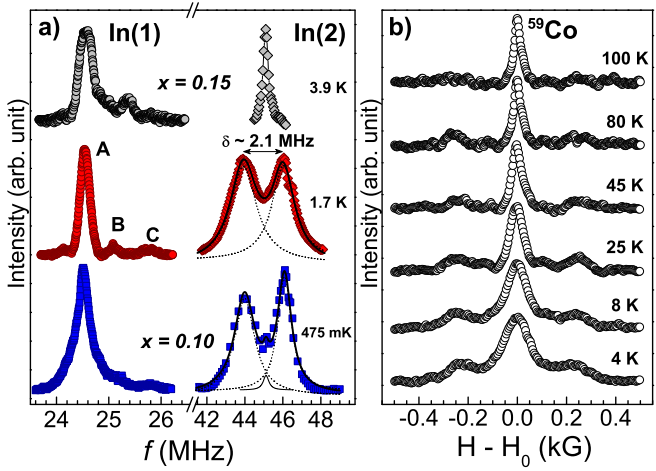}
\caption{\label{fig:spectra} (a) $^{115}$In NQR spectra  of the
In(1) and In(2) sites at 4 K ($x=0.15$, grey), at 1.5 K ($x=0.15$,
red) and at 0.45 K ($x=0.10$, blue). The A, B and C lines of the
In(1) are shown. The In(2) spectrum is split by the internal field
below $T_N$. Dotted lines are Gaussian fits, as described in the
text. (b) The central transition of the $^{59}$Co at $H_0 =33.824$
kG for the $x=0.15$ sample. The first satellite transitions are also
seen at $\pm220$ G.}
\end{figure}

All of measurements were made using crystals of \cecoincdx\ grown
from In flux as described in \cite{FiskCddoping}. Single crystal
were used with the field $\mathbf{H}\,||\,\hat{c}$ for the Knight
shift and linewidth measurements. \slrr\ was measured in zero field
using nuclear quadrupolar resonance (NQR) at the high symmetry In(1)
site in powdered samples, using conventional inversion recovery
pulse sequences \cite{CurroCeRhIn5}.  Below 1 K, the rf power was
reduced sufficiently to prevent heating during the time scale of the
measurement. The In NQR spectra are broadened relative to the pure
compound, and two new features emerge in the In(1) spectra, which we
term the B and C-sites (Fig. \ref{fig:spectra}a). The origin of
these sites is unclear, but we speculate that they are related to
nearest neighbor sites of the Cd dopants. The \slrr\ data reported
here were obtained at the A site, which we associate with the bulk
In(1) sites.

\begin{figure}
\includegraphics[width=\linewidth]{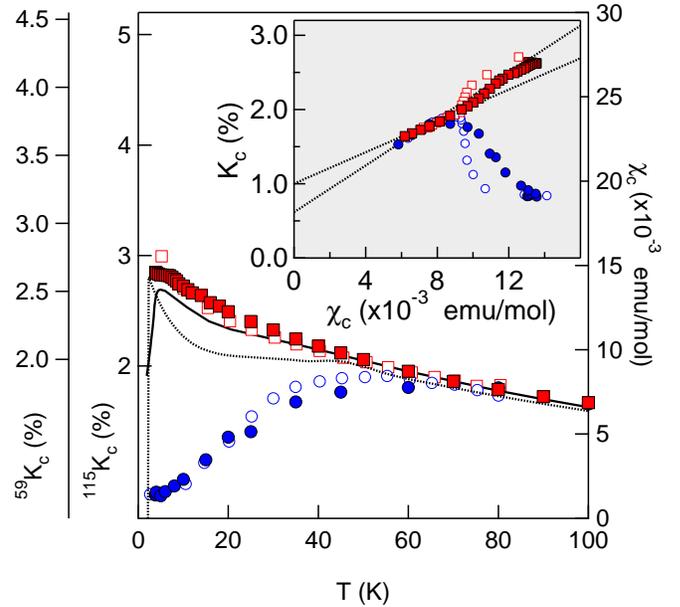}
 \caption{\label{fig:KvsChi} The Knight shift of the
$^{115}$In(1) ($\blacksquare$, $x=0.15$ and $\square$, $x=0$) and
$^{59}$Co ($\bullet$, $x=0.15$ and $\circ$, $x=0$) versus
temperature. The shift of the $x=0.10$ sample (not shown) is
qualitatively similar. The solid (dotted) line is the bulk
susceptibility for $x=0.15$ $(x=0)$ in the $c$ direction
\cite{shiftnote}. INSET: The Knight shifts versus the bulk
susceptibilities. The dotted lines are fits to the high temperature
data.} \vspace{-0.15in}
\end{figure}

In the \afc\ state, the In(1) NQR resonance moves down by $\sim 25$
kHz, whereas the In(2) resonance splits into two lines separated by
$\sim 2.1$ MHz (see Fig. \ref{fig:spectra}a). This response is
similar to that of the \cerhin, where the internal field at the
In(1) lies in the $ab$ plane and the internal field at the In(2)
lies along the $c$ axis \cite{CurroCeRhIn5}. The internal field at
the In(1) nearly vanishes even though the ordered moment remains
finite, which is analogous to the response of \cerhin\ under
pressure \cite{kitaokaCeRhIn5coexistence,AnnaCeRhIn5NSpressure}.
Recent neutron scattering experiments suggest that the ordering
wavevector in \cecoincdx\ is $\mathbf{Q} = (\pi/a, \pi/a, \pi/c)$
\cite{nicklasCdNS}. The splitting of the In(2) spectrum reveals
$H_{\rm int}(2) \approx 2.3$ kOe along the $c-$axis, nearly
identical to that observed at the same transition in \cerhin\ (2.7
kOe) and in the field-induced magnetic phase of pure \cecoin\
\cite{CurroCeRhIn5,CurroCeCoIn5FFLO}. Assuming the hyperfine
interaction is the same in both materials, this result suggests that
the ordered moment is $\sim 0.7$ $\mu_B$ in the \cecoincdx\ system.
Surprisingly, we find that the internal field is identical in both
the $x=0.10$ and $x=0.15$ materials, suggesting that the degrees of
freedom involved with the long range magnetic order are independent
of the superconducting degrees of freedom. The spectra also reveal
approximately 3\% of the sample volume with no internal field in the
$x=0.10$ sample (solid line in Fig. \ref{fig:spectra}). One
explanation for this result is that the sample is inhomogeneous, so
that some fraction of the sample sees magnetic order while the other
fraction sees superconductivity. However, if this were the case then
the jump in the specific heat at $T_c$ would be $\Delta C \sim 0.03
\gamma T_c \approx 13\; {\rm mJ}/{\rm mol\: K}$ (using $\gamma
\approx 350\: {\rm mJ}/{\rm mol\: K}^2$), about a factor of 6
smaller than observed \cite{FiskCddoping}. Therefore, the anomalous
3\% may be due to doping inhomogeneity on a macroscopic scale in the
powdered sample. The remaining 97\% of the sample volume shows
microscopic coexistence of antiferromagnetism and superconductivity.

\begin{figure}
\includegraphics[width=\linewidth]{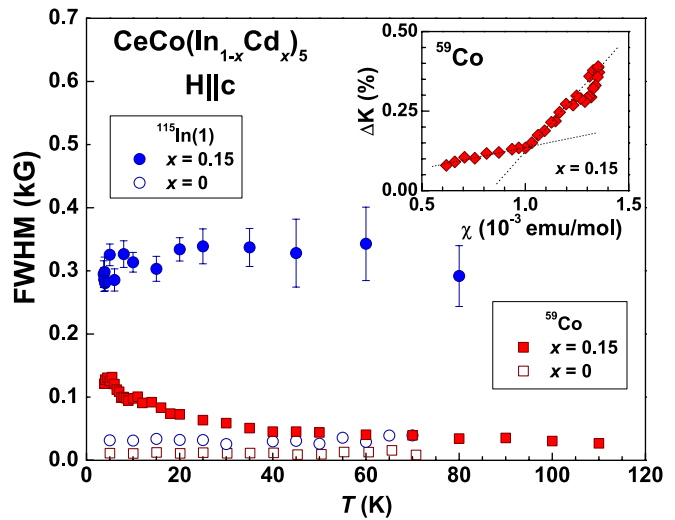}
\caption{\label{fig:linewidth} The linewidths (full width half
maxima) of the Co ($\square$, $x=0$ at 38 kOe, and $\blacksquare$,
$x=0.15$ at 33.824 kOe) and In(1) ($\circ$, $x=0$ at 38 kOe, and
$\bullet$, $x=0.15$ at 33.824 kOe) spectra. The linewidth of the
$x=0.10$ sample (not shown) is qualitatively similar. INSET: The Co
Knight shift distribution (FWHM/$\gamma H_0$) versus the bulk
susceptibility.} \vspace{-0.15in}
\end{figure}

Fig. \ref{fig:KvsChi} shows the temperature dependence of the Knight
shift of the In(1) A line and the Co \cite{shiftnote}. Like the pure
compound, there is a large Knight shift anomaly that develops below
a temperature $T^*\sim 60$ K \cite{CurroAnomalousShift}.  In fact,
both $T^*$ and the temperature dependence of $K$ remain almost
doping independent for both sites.  This result suggests that the
spin response at the bulk A sites remains unaffected by the Cd
doping, and that the difference in bulk susceptibility, $\chi_c(T)$,
may arise solely in the vicinity of the Cd.  This scenario is
further supported by the enhanced magnetic broadening as seen in
Fig. \ref{fig:linewidth}.  Doping broadens the In(1) NQR spectra by
a factor of ten due to the distribution of local electric field
gradients (EFGs). On the other hand, the EFG at the Co is nearly two
orders of magnitude smaller than that at the In(1) site, so it is
less sensitive to the distribution of local EFGs and more sensitive
to distributions of local magnetic fields. As observed at the
central transition
$\left(+\frac{1}{2}\leftrightarrow-\frac{1}{2}\right)$ (Fig.
\ref{fig:spectra}b and \ref{fig:linewidth}) the magnetic broadening
is significantly larger in the doped samples, and is strongly
temperature dependent.  The broadening is not directly proportional
to $\chi_c(T)$ but shows a change in behavior below 60 K (inset Fig.
\ref{fig:linewidth}). This result implies that the magnetic
broadening mechanism does not arise from a distribution of
demagnetization fields, but rather has microscopic origin, possibly
due to a variation in the local staggered magnetization in the
vicinity of the Cd \cite{bobroff}.

Fig. \ref{fig:T1vsT}b shows the temperature dependence of \slrr\ at
the In(1) site for several different samples.  In the disordered
state, \slrr\ is independent of doping above 5 K, and shows an
unusual $T^{1/4}$ dependence. This behavior contrasts with the
evolution of \slrr\ with pressure in pure \cecoin, \cerhin,
CeIn$_3$, and CeCu$_2$(Si,Ge)$_2$, where the magnitude of \slrr\ in
the disordered state changes dramatically as the ground state
evolves from antiferromagnetic to superconducting
\cite{kitaokacecoexistencereview,kohoriCeCoIn5pressure}.  Pressure
tuning modifies the Kondo interaction, $J$, and changes in the
ground state are reflected in \slrr\ through modifications of the
spin fluctuation spectrum. If the Cd doping induces changes to the
Fermi surface, modifying either $N(0)$, $J$, and hence the RKKY
interaction, then similar modifications will be evident in the spin
fluctuation spectra of the doped materials.  On the other hand, if
the Cd doping acts as a local seed for antiferromagnetic order, the
majority of the In sites will not sense any change in the spin
fluctuations until the correlation length, $\xi$, reaches a value on
the order of the dopant spacing ($\sim 4$ lattice spacings). Neutron
scattering studies suggest that $\xi$ is indeed of this order at
$T_N$ \cite{baoCeRhIn5INS}. This latter scenario offers an
explanation not only for the \slrr\ and $K$\ data, but also for the
sensitivity of the ground state to only a few percent of dopants.

Below $T_N$, \slrr\ in the $x=0.15$ sample agrees with that in
\cerhin\ down to 300 mK.  Surprisingly, \slrr\ remains constant down
to 45 mK.  This result suggests that even in the
antiferromagnetically ordered state, sufficient fluctuations remain
in some channel to relax the nuclei. It is not clear if these
fluctuations arise from excitations of the ordered magnetic
structure, possibly excited by interactions with the heavy
electrons; or from some other remaining degree of freedom, possibly
as a precursor to the onset of superconductivity at a lower
temperature.

In the $x=0.10$ sample, the onset of superconductivity is clearly
reflected in \slrr, which drops by more than an order of magnitude
below $T_c$.  However, the temperature dependence of \slrr\ below
$T_c$ is not simply $T^3$ as observed in pure \cecoin\
\cite{kohoriCeCoIn5QCP}.   We find that \slrr\ $\sim T^\alpha$ below
200 mK, where $\alpha=0.05(3)$ for $x=0.15$, $\alpha=0.25(3)$ for
$x=0.10$, and $\alpha=1.0(2)$ for $x=0$, so that
\slrr$(x,T\rightarrow0)$ increases with $x$. It is not clear if the
unusual sublinear behavior in the doped samples is related to the
normal state relaxation, or is part of a crossover to a constant at
lower temperature. In a d-wave superconductor \slrr\ $\sim T^3$ in
the clean limit, but impurity scattering can give rise to an
enhanced spin-lattice relaxation that varies linearly with
temperature for sufficiently low temperatures
\cite{CurroPuCoGa5,TouUPdAlT1,kitaokaCeRhIn5coexistence,zhengCeRhIrIn5,kitaokaCeRhIn5pressureGapless,kohoriURu2Si2,kitaokaCeCu2Si2}.
In superconductors with coexisting antiferromagnetism, the low
energy density of states is even more sensitive to the presence of
impurity scattering \cite{bang}. The Cd dopants in \cecoincdx\
certainly provide a source for impurity scattering, but the
sublinear behavior cannot be explained by conventional in-gap
states.  Indeed, if the Cd served as a pair-breaking impurity, then
superconductivity could not be recovered by applying pressure
\cite{FiskCddoping}. One possibility is that the density of states
has an unusual energy dependence, which may arise as a result of
interactions between the nodal quasiparticles and the low energy
excitations of the antiferromagnetic structure.
CeRh$_{0.5}$Ir$_{0.5}$In$_5$ has coexisting antiferromagnetic and
superconducting order, and both In sites show enhanced \slrr\ in the
superconducting state. However, the enhanced relaxation turns on at
the In(2) site at a higher temperature than for the In(1)
\cite{zhengCeRhIrIn5}.  If the origin of the enhanced \slrr\ were
purely from impurity scattering, then the onset temperature should
not depend on the particular site. This result suggests instead that
the enhanced \slrr\ is dominated by fluctuations of the
antiferromagnetic order since the static hyperfine field at the
In(2) is larger than that at the In(1). It is possible that the
unusual temperature dependence of the enhanced relaxation in
\cecoincdten\ is driven by such a mechanism, but it is not clear
what the temperature dependence should be in this case.

In summary we have measured the NMR response of the \cecoincdx\ and
found evidence that Cd doping does not modify the Fermi surface, but
rather nucleates local antiferromagnetic droplets. When the droplets
overlap the system undergoes long range order and the
antiferromagnetism and superconductivity coexist microscopically. We
thank M. Graf, T. Park, R. Movshovich, D. Pines, and F. Ronning for
enlightening discussions. LDP and ZF acknowledge support from
NSF-DMR-0533560 and BY thanks support from the MOE ATU Program. Work
at Los Alamos National Laboratory was performed under the auspices
of the U.S. Department of Energy.


\begin{thebibliography}{27}
\bibitem[{shi()}]{shiftnote}
\bibinfo{note}{The Co data in the $x=0.15$ sample have been shifted vertically
  by 0.15 \% to agree with the data in the pure compound above 60 K. The
  difference between the two measurements is likely due to changes in the field
  calibration of the magnet.}
\end{thebibliography}

\end{document}